Denoising Improves Cross-Scanner and Cross-Protocol Test-Retest Reproducibility of Higher-Order Diffusion Metrics

Benjamin Ades-Aron[a,b,*], Santiago Coelho[a], Gregory Lemberskiy[b], Jelle Veraart[a], Steven Baete[a], Timothy M. Shepherd[a], Dmitry S. Novikov[a], Els Fieremans[a]

[a]Bernard and Irene Schwartz Center for Biomedical Imaging, Department of Radiology, New York University School of Medicine, New York, NY, USA

[b]Microstructure Imaging Inc., Brooklyn, NY, USA

Corresponding author: Benjamin Ades-Aron

Email address: benjamin.ades-aron@nyulangone.org

Phone: (516) 965-0756

Acknowledgements and Disclosures: The authors would like to thank Genevieve Barroll and the research support staff at the Center for Biomedical Imaging who were instrumental in aiding the recruitment and data collection process for this study. This work was performed under the rubric of the Center for Advanced Imaging Innovation and Research (CAI2R, https://www.cai2r.net), a NIBIB Biomedical Technology Resource Center (NIH P41-EB017183). This work has been supported by NIH under NINDS award R01 NS088040 and NIBIB awards R01 EB027075 and EB028774.

Denoising Improves Cross-Scanner and Cross-Protocol Test-Retest Reproducibility of Higher-Order Diffusion Metrics

Abstract

*Objectives:* The clinical translation of diffusion MRI (dMRI)-derived quantitative contrasts hinges on robust reproducibility, minimizing both same-scanner and cross-scanner variability. As multi-site data sets, including multi-shell dMRI, expand in scope, enhancing reproducibility across variable MRI systems and MRI protocols becomes crucial. This study evaluates the reproducibility of higher-order diffusion metrics (beyond conventional diffusion tensor imaging), at the voxel and region-of-interest (ROI) levels on magnitude and complex-valued dMRI data, using denoising with and without harmonization.

*Materials and Methods:* We compared same-scanner, cross-scanner, and cross-protocol variability for a standardized multi-shell dMRI protocol (2-mm isotropic resolution, $b$=0, 1000, 2000 s/mm$^2$) in 20 subjects. We first evaluated the effectiveness of Marchenko-Pastur Principal Component Analysis (MPPCA) based denoising strategies for both magnitude and complex data to mitigate noise-induced bias and variance, to improve dMRI parametric maps and reproducibility. Next, we examined the impact of denoising under different population analysis approaches, specifically comparing voxel-wise versus region of interest (ROI)-based methods. We also evaluated the role of denoising when harmonizing dMRI across scanners and protocols.

*Results:* The results indicate that DTI and DKI maps visually improve after MPPCA denoising, with noticeably fewer outlier in kurtosis maps. Denoising, either using magnitude or complex dMRI, enhances voxel-wise reproducibility, with test-retest variability of kurtosis indices reduced from 15-20% without denoising to 5-10% with denoising. Complex dMRI denoising reduces the noise floor by up to 60%. In ROI-analyses, denoising not only reduced variability across scans and protocols, but also increased statistical power, with reduction in sample size requirements by up to 40% for detecting differences in mean diffusivity across populations. Combining denoising with linear-RISH harmonization, in voxel-wise assessments, improved intra-scanner test-retest intraclass correlation coefficients (ICCs) for FA from moderate (0.62 Prisma, 0.24 Skyra) to excellent repeatability (0.93 Prisma, 0.83 Skyra) over harmonization alone.

*Conclusions:* MPPCA denoising, either over magnitude or complex dMRI data, enhances the reproducibility and precision of higher-order diffusion metrics across same-scanner, cross-scanner, and cross-protocol assessments. As we demonstrate, the enhancement in data quality and precision facilitates the broader application and acceptance of these advanced imaging techniques in both clinical practice and large-scale neuroimaging studies.

*Abbreviations:*

Marchenko-Pastur Principal Component Analysis (MPPCA), diffusion Magnetic Resonance Imaging (dMRI), Diffusion Tensor Imaging (DTI), Diffusion Kurtosis Imaging (DKI)*,* Alzheimer's Disease Neuroimaging Initiative (ADNI), Adolescent Brain Cognitive Development (ABCD), Human Connectome Project (HCP), Concordance correlation Coefficient (CCC), Intraclass Correlation Coefficient (ICC), Region of Interest (ROI), Coefficient of Variation (CV).

1. Introduction

Diffusion magnetic resonance imaging (dMRI) is a noninvasive method enabling clinicians and researchers to characterize tissue microstructure beyond the nominal resolution of MRI. It measures the restricted diffusion of water within tissues[1], providing insights into neurological structure and function[6-9]. Recent large scale data initiatives like the Alzheimer's Disease Neuroimaging Initiative (ADNI)[2], Adolescent Brain Cognitive Development (ABCD)[3], the Human Connectome Project (HCP)[4], and UK Biobank[5], aim to utilize multi-shell dMRI to achieve a more comprehensive characterization of brain microstructure than possible with conventional diffusion tensor imaging (DTI). Higher-order diffusion signal representations and biophysical models[6–9] have been shown to characterize age related tissue changes in healthy populations[9,10] and pathology stemming from disease[12–16].

Despite its potential, the variability of dMRI-derived parameters, across different scanners and even on the same scanner, challenges its translation into clinical practice.[73]. This variability stems from factors like changing NMR contrast (e.g., due to operating at different echo time or field strength), imaging artifacts like bias due to motion or Gibbs ringing, and MRI noise which introduces random fluctuations and bias from the non-central-$\chi$ MRI noise floor[17,18]. These artifacts, coupled with poor noise propagation through the high-order representations (such as diffusion kurtosis imaging, DKI), lead to large outliers in quantitative maps[66], further reducing the reproducibility of such parameters. These variabilities limit MRI to a qualitative rather than a quantitative modality, particularly as MRI technology itself rapidly advances, prompting frequent updates to study protocols (such as parallel imaging[19,20] and simultaneous multislice (SMS) imaging[21]) that affect NMR contrast and noise floors. Low Signal-to-Noise-Ratio (SNR) reduces the precision of dMRI and amplifies scanner- and protocol-dependent biases[26], necessitating effective noise suppression strategies[79,80] to enhance data reliability, harmonization, and enable more accurate comparisons of diffusion parameters between healthy and pathological tissues.

In this study we evaluate two noise reduction techniques based on Random Matrix Theory – MPPCA for both magnitude and complex image data - to reduce within-scanner, cross-scanner, and cross-protocol variability. We compare repeatability and reproducibility of three kinds of processing: (i) without denoising; (ii) using state-of-the-art Marchenko Pastur Principal Component Analysis (MPPCA) denoising[27] for magnitude images; and (iii) adopting MPPCA to denoise complex images[28] (to maximize the precision of conventional DTI and higher-order diffusion metrics). MPPCA identifies an optimal threshold for noise-only singular values using the Marchenko-Pastur distribution. This technique and its variations have become a widely used component of dMRI pre-processing pipelines[29–31]. Complex MPPCA operates on complex-valued data (after coil combination), and is enabled through phase estimation and unwinding prior to applying the MPPCA algorithm[29,32].

We acquired five dMRI datasets in 20 human subjects. For each subject, we acquired two measurements using the same dMRI protocol (scan-rescan) on the same scanner, and an additional third dMRI measurement on the same scanner with modified acquisition parameters resulting in an overall lower SNR. The latter dMRI protocol was also measured twice (scan-rescan) on a second scanner (with weaker gradients, resulting in a longer echo time, and less extensive dMRI protocol). Both dMRI protocols were optimized to have the highest possible SNR at a fixed resolution and diffusion scheme for the specific scanner (Table 1). Our results indicate that MPPCA-based denoising significantly reduces bias and variance caused by differences in SNR related to hardware and imaging protocol and improves the accuracy of statistical comparisons across scanners and dMRI protocols.

2. Methods

*2.1. Clinical MRI Protocol*

This HIPAA-compliant prospective study was approved by the local institutional review board. After providing informed consent, 20 healthy volunteers (10 male / 10 female, age = 32.2 ± 9.7 years) underwent brain dMRI on Siemens Magnetom Prisma and Skyra 3T systems. All protocols included a diffusion weighted monopolar spin echo EPI sequence with 2-mm isotropic resolution, using a 20-channel head coil and *b*-values commonly used in multi-shell acquisitions[4,5] including $b$ = 0, 1000, and 2000 s/mm$^2$ shells, as detailed in Table 1. Acceleration was performed using 6/8 partial Fourier, in-plane parallel imaging (GRAPPA factor 2), and simultaneous multislice (SMS) acceleration with multiband factor 2. Raw data from all acquisitions were saved in the Siemens TWIX data format and fed into a standard reconstruction pipeline implemented in MATLAB for the purpose of saving phase maps. The pipeline included Nyquist ghost correction[34], coil noise decorrelation[35], Projection onto Complex Sets (POCS)[36], slice-wise GeneRalized Autocalibrating partially parallel acquisitions (GRAPPA)[37,20], Controlled Aliasing In Parallel Imaging Results IN Higher Acceleration (CAIPIRINHA) shift[38], in-plane GRAPPA, trapezoidal regridding[39], and adaptive combine[40]. During this reconstruction, both magnitude and phase images were saved for use during subsequent denoising and artifact correction steps.

Five test-retest dMRI datasets were acquired for each subject, as listed in Table 1. Two test-retest datasets were acquired on the Prisma system using the dMRI protocol with the shortest possible TE = 92 ms. A third dataset was acquired on the Prisma using a modified protocol (with TE = 127 ms) to study cross-protocol reproducibility. Next, two datasets were acquired on the Skyra system using the dMRI protocol with the shortest possible TE = 127 ms (due to smaller maximal gradient). Scan time for all dMRI protocols (subset used here) was approximately the same at 7.5 minutes, with respectively 60 and 50 directions on the Prisma and Skyra systems.

| Scanner | number of datasets | dataset labels | TR (ms) | TE (ms) | $b = 0$ s/mm$^2$ | $b = 1000$ s/mm$^2$ | $b = 2000$ s/mm$^2$ | acq. time min:sec | gradient (mT/m) |
|---|---|---|---|---|---|---|---|---|---|
| Prisma | 2 | $P_{92}^{(1,2)}$ | 5300 | 92 | 5 | 20 | 40 | 7:28 | 80 |
| Prisma | 1 | $P_{127}^{(1)}$ | 6700 | 127 | 5 | 15 | 30 | 7:28 | 80 |
| Skyra | 2 | $S_{127}^{(1,2)}$ | 6700 | 127 | 5 | 15 | 30 | 7:27 | 40 |

Table 1: The three dMRI protocols used during this study. Each of the 20 subjects underwent these three protocols to acquire in total five datasets, including a test-retest dataset on the Prisma, test-retest on the Skyra, and a single dataset acquired on the Prisma with protocol matched to the Skyra.

Between test-retest acquisitions, subjects were removed and then placed back into the scanner. Each test-retest dataset included a reverse phase encoding $b = 0$ image to correct for EPI-induced distortions. A 1-mm isotropic T1-weighted MP-RAGE (Magnetization Prepared Rapid Gradient Echo) sequence (TR/TE/TI = 2200/3.17/900 ms) was acquired on the Prisma system for co-registration of all 5 datasets per subject.

These five datasets were used to perform three comparisons.

1. Effect of denoising on *within-scanner variability (repeatability)* by comparing Prisma$^{(1)}$ vs rescan Prisma$^{(2)}$ with TR/TE 5300/92ms ($P^{(1)}_{92}$ vs $P^{(2)}_{92}$) and comparing Skyra$^{(1)}$ vs rescan Skyra$^{(2)}$ with TR/TE=6700/127ms, reffered to as $S_{127}^{(1)}$ vs $S_{127}^{(2)}$.

2. Effect of denoising on *cross-protocol variability (reproducibility)* by comparing data from the same scanner with unmatched TE and SNR: Prisma[(1)] TR/TE=5300/92ms vs Prisma[(2)] TR/TE=6700/127ms, referred to as $P_{92}^{(1)}$ vs $P_{92}^{(2)}$.

3. Effect of denoising on *inter-scanner variability (reproducibility)* by comparing data from different scanners, but matched TE: Skyra[(1)] TR/TE=6700/127ms vs Prisma TR/TE=6700/127ms, referred to as $S_{127}^{(1)}$ vs $P_{127}^{(1)}$.

*2.2. MPPCA-based Denoising methods*

Denoising was performed using an augmentation of the MPPCA algorithm[27,41]. MPPCA exploits data redundancy in the singular value decomposition (SVD) / principal component analysis (PCA) domain using properties of the eigenspectrum of random covariance matrices, which yields an objective number $p$ of signal-carrying components to be kept and all other components removed as purely noise-carrying.

A data matrix $X = \{X_{mn}\} \equiv \{S_m(\boldsymbol{r_n})\}$ of size $M \times N$ is formed by $M$ measurements from $N$ voxels in a patch. Let $M' = \min(M, N)$ and $N' = \max(M, N)$. Low-rank denoising corresponds to keeping $p$ largest components in the SVD of $X = \sum_{i=1}^{M'} s_i u_i \otimes v_i$, where $u_i$ and $v_i$ are the left and right singular vectors. Equivalently, PCA corresponds to keeping $p$ top eigenvalues $x_i = s_i^2$ explaining most of the variance of the (sample) covariance matrix $XX^\top$. MPPCA yields the number $p$ of the components of $X$ to keep, by identifying the Marchenko-Pastur (MP) distribution formed by the remaining $M' - p$ components that correspond to pure noise in the limit $M, N \gg 1$ and $p \ll M, N$ (low-rank condition). MPPCA self-consistently finds $p$, and noise variance $\sigma^2$ as the sum over the $M' - p$ components attributed to the MP distribution.

In recent years there having been several proposed additions to the original MPPCA algorithm, including symmetric thresholding of singular values[31], eigenvalue shrinkage[42], advanced patching methods including non-local spatial matching[43–46], angular matching[47], and structural adaptation[48]. Here, MPPCA is augmented in three stages: (i) Adaptive patching to select the signals forming a data matrix $X$ around a given voxel; (ii) Symmetric SVD threshold selection[31]; (iii) Singular value shrinkage[42]. We here describe each stage:

 i. *adaptive patching.* — In contrast to the original MPPCA approach where patches are squares or cubes around a given voxel (e.g., 5 × 5 × 5 voxels), we enhanced the spatial redundancy, in addition to the redundancy in the measurements (here in the diffusion *q*-space), by pre-selecting voxels that have similar ground truth. Our goal is to minimize the number $p$ of components, and to maximize their contributions $s_i$, such that they describe most of the signal — this is the assumption of the underlying noise-free signal to be of low-rank. The ultimate best choice of the patch would be having all $N$ voxels with the same ground truth; in this case, $p \equiv 1$ (the rank of noise-free $XX^\top$ is 1), irrespective of the complexity of their signal $S_m$. The desire to have a few large $s_i$, $i = 1, \ldots, p$, (as opposed to many smaller ones) comes from the fact that components whose noise-free singular values $s_i^{(0)} < s_* = \sigma(MN)^{1/4}$ are below the threshold $s_*$ become indistinguishable from noise. They fall below what's referred to as the *phase transition*[49,50]. By selecting a set of voxels that have similar underlying tissue priors, we intend to overcome the information loss associated with many components potentially falling under the phase transition threshold $s_*$, if all voxels in a patch were to have fairly different signals.

Hence, for a voxel at $\boldsymbol{r_0}$, we would like to include the signals $S_m(\boldsymbol{r_n})$ such that both $S_m$ are close to each other for different $\boldsymbol{r_n}$, and the voxels $\boldsymbol{r_n}$ are not too far from $\boldsymbol{r_0}$. Formally, we introduce the "distance" between signals

$$w_{\alpha,\beta}[S(\boldsymbol{r}), S(\boldsymbol{r'})] = |\boldsymbol{r} - \boldsymbol{r'}|^\alpha \cdot |S(\boldsymbol{r}) - S(\boldsymbol{r'})|^\beta. \qquad (1)$$

Here $|\boldsymbol{r} - \boldsymbol{r'}|$ is the Euclidean distance between voxels in 3 dimensions, and $|S(\boldsymbol{r}) - S(\boldsymbol{r'})| = \sqrt{\sum_m |S_m(\boldsymbol{r}) - S_m(\boldsymbol{r'})|^2}$ is the Euclidean distance between signals (the norm over the measurement index *m*). The balance of preferring the distance between voxels and between signals is tuned by the choice of exponents α and β. Here we choose α = 1 and β = 2 based on an empirical observation of improved denoising performance when β > α (preferring similarity of signals to the distance from $\boldsymbol{r_0}$). When α ≫ β this method converges to the original MPPCA patching implementation (local signal-independent patch around $\boldsymbol{r_0}$). Based on the above distance function, we choose

a patch around voxel at $r_0$ as $N$ voxels (including $r_0$), for which $w_{1,2}[S(r_n), S(r_0)]$ is the smallest; here $N$ was fixed to 100, Fig. 1.

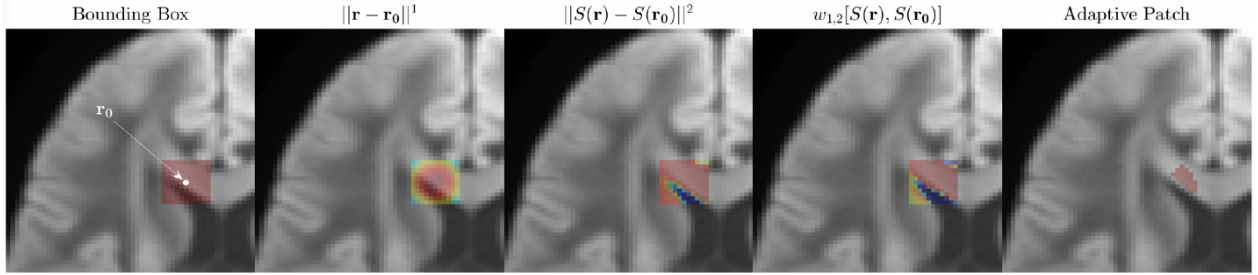

Figure 1: Example of how distances $w_{\alpha,\beta}[S(r_n), S(r_0)]$ Eq. (1), are generated for adaptive patching. These distances are used to threshold a cubic patch into the shape of local anatomy. We first choose a sufficiently large bounding box (in this example an 11 × 11 × 11 bounding box was chosen for demonstration purposes, in the study a 7×7×7 bounding box was used). Next, the Euclidean distances between voxels (2nd panel) and signals (3rd panel) are calculated. The weights (1) are computed based on the product of the two distances, tuned by exponents α and β. The patch selection corresponding to $N$ = 100 is shown in the last panel.

ii. *Symmetric thresholding.* — For the pure-noise case of $p = 0$, all components of $XX^\intercal$ form the MP distribution, which has two independent properties: $\sigma^2 = \sum x_i/(MN)$ and $\sigma^2 = (x_1 - x_{M'})/4\sqrt{MN}$. MPPCA uses these two properties to define two functions $\sigma_{1,2}(p)$ accounting for the noise variance from the bottom $M' - p$ components, with $p$ being the solution for $\sigma_1(p) = \sigma_2(p)$. Here, we use the following definitions[29,31]:

$$\sigma_1^2(p) = \frac{1}{(N'-p)(M'-p)} \sum_{i=p+1}^{M'} x_i, \tag{2.1}$$

$$\sigma_2^2(p) = \frac{x_{p+1} - x_{M'}}{4\sqrt{(N'-p)(M'-p)}} \tag{2.2}$$

These are symmetric in $M'$ and $N'$, whereas the original MPPCA formulation had $N'$ instead of $N' - p$ in Eqs. (2). This symmetric augmentation of $\sigma_{1,2}(p)$ empirically provides a more robust estimation of MP threshold for not very large $N$ and $M$, and works well for $M \approx N$. It is also practically important that the patch size is allowed to vary, with possibilities of both $M > N$ and $M < N$. In this study $N$ was fixed at 100, and $M$ was the total number of diffusion measurements for all subjects and scan-rescan datasets. $M = 65$ for $P_{92}^{(1,2)}$, $M = 50$ for $S_{127}^{(1,2)}$, and $M = 50$ for $P_{127}^{(1)}$.

iii. *Singular value shrinkage.* — To overcome the eigenvalue repulsion due to noisy components, we reduce the sample singular values $s_i$ when recombining selected principal components into a low-rank matrix $\widehat{X_\eta} =$ $\sqrt{M'}\sigma \sum_{i=1}^{p} \eta\left(s_i/(\sqrt{M'}\sigma)\right) u_i \otimes v_i$. According to Gavish and Donoho (42), the optimal shrinker function based on minimizing MSE of the Frobenius norm can be expressed as

$$\eta(s) = \begin{cases} \frac{1}{s}\sqrt{(s^2 - \gamma - 1)^2 - 4\gamma}, & s > 1 + \sqrt{\gamma} \\ 0, & s \leq 1 + \sqrt{\gamma} \end{cases} \quad (3)$$

where $\gamma = M'/N' \leq 1$.

### 2.3. Denoising experiments

dMRI raw data was preprocessed in three different ways for comparison:

i. Magnitude (non-denoised) dMRI data.

ii. MPPCA: Magnitude dMRI data were MPPCA-denoised (including the adaptive patch, symmetric thresholding, eigenvalue shrinkage).

iii. MP-Complex: dMRI phases were first denoised using MPPCA (symmetric thresholding) using a 15×15 2d box-shaped kernel within each slice. The denoised and spatially smoothly varying phase $\phi_{dn}(r)$ is then unwound according to: $S_{real} = \text{Re}(S_{complex} e^{-i\phi_{dn}})$. Finally, the noisy phase-unwound signal $S_{real}$ is denoised using an adaptive 3d moving patch, symmetric thresholding, and eigenvalue shrinkage. Phase unwinding helps remove spurious components arising due to shot-to-shot phase variations in dMRI[28,32]. In addition to improved denoising performance (due to the two-pass approach), MP-Complex also reduces the Rician noise floor of the data, reducing bias and increasing the precision of downstream parameter estimation.

### 2.4. Processing Pipeline and Diffusion Parameter Estimation

All dMRI images were processed using the DESIGNER pipeline[51,78], using Partial Fourier induced Gibbs artifact correction[52,83], EPI distortion correction[53], and eddy-current and motion correction (including slice-wise outlier replacement)[54,55]. The signal's normalized rotational invariants of zeroth ($S_0$, also known as spherical mean) and second order ($S_2$) were linearly estimated for each diffusion shell[56] to provide a convenient basis for population-wise registration.

Diffusion and kurtosis tensors were estimated through unconstrained weighted linear least squares[57] from which we derived mean diffusivity (MD), axial diffusivity (AD), radial diffusivity (RD), fractional anisotropy (FA), mean kurtosis (MK), axial kurtosis (AK), and radial kurtosis (RK).

The minimal protocol used in this study uses kurtosis estimation from the cumulant expansion up to $\mathcal{O}(b^2)$:

$$\ln \frac{S(b,\hat{\boldsymbol{n}})}{S|_{b=0}} = -bD(\hat{\boldsymbol{n}}) + \frac{1}{6} b^2 \bar{D}^2 W(\hat{\boldsymbol{n}}) + \cdots \qquad (4)$$

where $D(\hat{\boldsymbol{n}}) = D_{ij} n_i n_j$, $W(\hat{\boldsymbol{n}}) = W_{ijkl} n_i n_j n_k n_l$, and

$$\mathrm{MD} \equiv \bar{D} = \int_{|\hat{\boldsymbol{n}}|=1} \frac{d\hat{\boldsymbol{n}}}{4\pi} D(\hat{\boldsymbol{n}}) = \frac{1}{3} D_{ii} \qquad (5)$$

(the Einstein's convention of summation over pairs of repeated indices is assumed throughout).

The definition of mean kurtosis is ambiguous in the dMRI literature. The original paper of Jensen et al. (24) suggested that MK is an angular average of the directional kurtosis $K(\hat{\boldsymbol{n}}) = \bar{D}^2 W(\hat{\boldsymbol{n}})/D^2(\hat{\boldsymbol{n}})$:

$$\mathrm{MK} = \bar{D}^2 \int_{|\hat{\boldsymbol{n}}|=1} \frac{d\hat{\boldsymbol{n}}}{4\pi} \frac{W(\hat{\boldsymbol{n}})}{D^2(\hat{\boldsymbol{n}})}. \qquad (6)$$

While perhaps more intuitive, this definition suffers from two drawbacks. First, fundamentally, the result cannot be compactly represented as a trace of a certain tensor (e.g., analogously to Eq. (5)) — due to a nontrivial directional dependence of the denominator. Second, practically, this definition leads to a relatively low precision and is strongly affected by outliers, which come from directions where diffusivity $D(\hat{\boldsymbol{n}}) \ll \bar{D}$ is small. An alternative definition involves angular average of only $W(\hat{\boldsymbol{n}})$:[58]

$$\mathrm{MW} \equiv \bar{W} = \int_{|\hat{\boldsymbol{n}}|=1} \frac{d\hat{\boldsymbol{n}}}{4\pi} W(\hat{\boldsymbol{n}}) = \frac{1}{5} W_{iikk}. \qquad (7)$$

The last equation follows, e.g., from Appendix C of (56). Hence, instead of performing an integral over a limited number of directions, we can calculate $\bar{W}$ very fast and exactly from the estimated $W$ tensor by simply taking a full trace. This quantity is dimensionless (as is MK), and has qualitatively similar contrast to MK. The axial and radial kurtoses, AW and RW, are calculated as projections onto the principal fiber direction and onto the plane transverse to it, respectively:

$$\begin{aligned}
AD &= D(\widehat{\boldsymbol{v}}_1) \\
RD &= \int_{|\widehat{\boldsymbol{n}}|=1} \frac{d\widehat{\boldsymbol{n}}}{2\pi} D(\widehat{\boldsymbol{n}})\delta(\widehat{\boldsymbol{n}} \cdot \widehat{\boldsymbol{v}}_1) \\
AW &= W(\widehat{\boldsymbol{v}}_1) \\
RW &= \int_{|\widehat{\boldsymbol{n}}|=1} \frac{d\widehat{\boldsymbol{n}}}{2\pi} W(\widehat{\boldsymbol{n}})\delta(\widehat{\boldsymbol{n}} \cdot \widehat{\boldsymbol{v}}_1)
\end{aligned} \qquad (8)$$

While MK, AK and RK are more commonly used in literature than MW, AW and RW, both parameter definitions will be generated in what follows, owing to the lower probability for outliers when using $W$ compared to $K$ (as shown in Fig. 3 in Section 3 below).

*2.5. Data Analysis and Statistics*

To evaluate the effect size of the denoising step in preprocessing on any dMRI parameter $x$ used in this study, we assess the variability across pairs of scans by the coefficient of variation $CV_x = \sigma_x/\mu_x$, where the estimate of the mean is $\widehat{\mu_x} = \frac{1}{2}(x_1 + x_2)$, and the estimate of variance $\widehat{\sigma_x} = \frac{\sqrt{\pi}}{2}|x_1 - x_2|$. Here $x_{1,2}$ refer to the parameter estimates from the two scans being compared. Test-retest variability was compared within scanner, across scanners, and across echo times, on a region of interest (ROI) level, and using a voxel-wise approach. In addition, concordance analysis and the concordance correlation coefficient (CCC)[59] were used to assess reproducibility and agreement between scans.

*Voxel-wise analysis*: For each subject, a multimodal template was generated based on $S_0$ rotational invariant maps and FA maps from each of the five repeated acquisitions. The subject-wise template was generated with Mrtrix3[61], using rigid registrations to a midpoint space, resulting in template maps for each of the 20 subjects. Next, a population template was generated based on all 20-subject template $S_0$ and FA pairs using nonlinear deformations (again using Mrtrix3). Rigid transformations from original space and warps to population space were concatenated, and then parametric maps for each subject, scan, and denoising level were transformed to the population template using cubic spline interpolation. For voxel-wise analyses, coefficients of variation were computed in the space of subject templates (after transforming parameters using rigid transforms) and in the space of the population template (by transforming parameters using nonlinear transforms). Nonlinear transforms have a small confounding effect on CV (due to interpolations performed on parametric maps) compared to rigid, therefore CVs computed after the nonlinear warp were only used to create figure 4. CV between test-retest data was computed after averaging parametric maps over all 20 subjects in the unbiased population midpoint space. We performed a concordance analysis showing the degree of concordance correlation between test-retest pairs in pooled voxels over all subjects and show Bland-Altmann plots to measure the error between the same pairs.

*Harmonization comparison:* The effect of harmonization was compared to denoising for data with bias due to varying echo times (and noise floors). Harmonization was performed using the state-of-the-art linear-RISH[82] method. Linear-RISH uses mean rotationally invariant spherical harmonic (RISH) features over a population to normalize those of each individual, and after normalization, comparisons were performed in subject's native space. All transformations were computed once and applied to all datasets to eliminate registration induced error.

We used the intra-class correlation coefficient (ICC) to measure voxel-wise one-to-one agreement in white matter, we compared harmonization without denoising, denoising alone, and using both harmonization and denoising.

*ROI analyses* were performed in the original space of each scan (to minimize registration bias in CV measurements) by computing a nonlinear registration between the JHU white matter (WM) atlas[61] and the overall population template, and applying the inverse warps used to transform each subject map to the population average. Analyzed WM ROIs include genu, body, and splenium of the corpus callosum, internal capsule, and corona radiata. Since the original JHU ROIs[61] are quite large, we thresholded them in template space (not in subject space) by removing voxels in lowest 5th percentile of FA to minimize misregistration and partial-volume effects. All these ROIs were combined into a single large WM region when tabulating full WM scale statistics. Gray matter (GM) ROIs were derived using Freesurfer[62], and CVs were computed in the thalamus. Freesurfer was performed using the MP-RAGE acquired for each subject as an input, and ROIs were rigidly registered to diffusion space using the preprocessed b0 image as a registration reference.

A comparison of how denoising impacts the noise floor in diffusion weighted images was performed to measure the level of bias reduction that comes from unwinding the denoised phase in complex data. We measured the noise floor by computing the spherical mean signal over directions $\bar{S}|_{b=2000}$ in the ventricles at $b = 2000 s/mm^2$. Due to diffusion attenuation, there is negligible signal ($\sim e^{-6} \approx 0.002$) left in CSF at this gradient strength, therefore $\bar{S}|_{b=2000}/\sqrt{\pi/2}$ can serve as a noise level estimator for a Rayleigh-distributed random variable. Ventricular segmentation was performed using Freesurfer as described above. The noise floor was further normalized by the signal at $b = 0$:

$$\text{Noisefloor} \equiv \frac{\sigma}{S_0} = \frac{1}{\sqrt{\pi/2}} \frac{\bar{S}|_{b=2000}}{S|_{b=0}}. \tag{9}$$

Finally, a power analysis was performed to test the number of subjects necessary to detect a 10% effect size based on group means and standard deviations observed in this healthy control population. A 10% effect size was chosen based on Cohen's-d values reported in literature[63], where for various psychiatric disorders groupwise differences in mean FA were found up to 42%. We used the formula:

$$n_1 = \frac{2(Z_{\alpha/2} + Z_\beta)^2 \sigma_p^2}{(\mu_{x1} - \mu_{x2})^2} \tag{10}$$

Here $n_1$ is the sample size for a single group, $\mu_{1,2}$ refers to the means of the two groups being compared and $\sigma_p$ is the standard deviation of a parameter $x$ over the clinical population ($\sigma_x$ described previously is a subset of the noise in $\sigma_p$). $Z_\alpha$ is the z-score for type one error level α, and $Z_\beta$ for type two error β[64]. This simulated the sample size required to reach voxel-wise statistical significance in a in a two-sided t-test with 80% power and α = 0.05. Noise level was selected based on results from voxel-wise CV given in Table 2. We created two groups whose diffusion and kurtosis parameters differed by 10% and measured the number of subjects required to reach statistical significance.

3. Results

Figure 2 shows representative fully processed diffusion and kurtosis maps for a single subject (26-year-old female) for test-retest data acquired on the Prisma scanner for MD, MK, MW, and FA. Two effects are notable from these maps: 1) The improved qualitative denoising effect from no denoising, to magnitude, to complex approaches. 2) The large decrease in kurtosis outliers observable as "black voxels" in MK maps both due to denoising, and by moving to the MW representation.

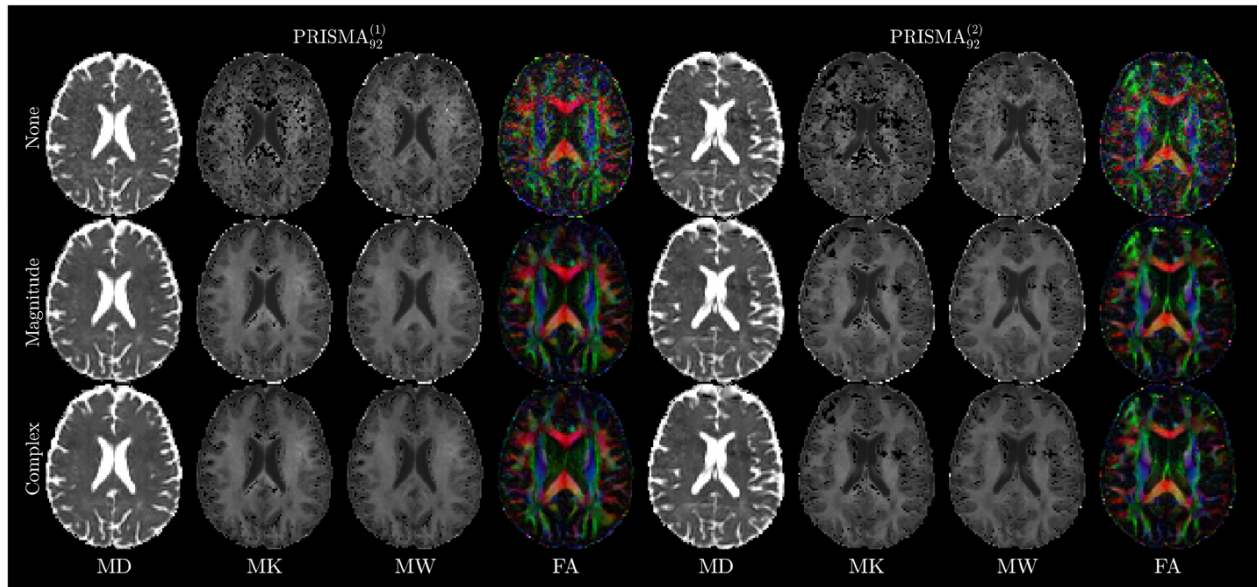

Figure 2: Illustration of different denoising methods on DTI and DKI measures (MD, MK, MW and FA) across scan-rescan data on the Prisma for a representative subject (26-year-old female). None: non-denoised data, Magnitude: MPPCA denoised magnitude data, Complex: MPPCA denoised complex data. Rician bias reduction is evident in the increased contrast of FA images in complex denoised data. Outliers were reduced in kurtosis maps both through denoising and the use of $W$ representation.

The number of kurtosis outliers present was quantified by counting the number of voxels $< -1$ in RK and RW maps. Figure 3 shows the strong reduction in outliers over the internal capsule (single subject) in mean kurtosis using each denoising method. MK measured on the Skyra system, has almost 70% of voxels labelled as outliers without denoising. By denoising and using MW instead only 2% of tissue is classified as outliers, most reduction coming from switching to $W$ representation (10× reduction in outliers), compared to denoising ($2 - 3$× reduction in outliers).

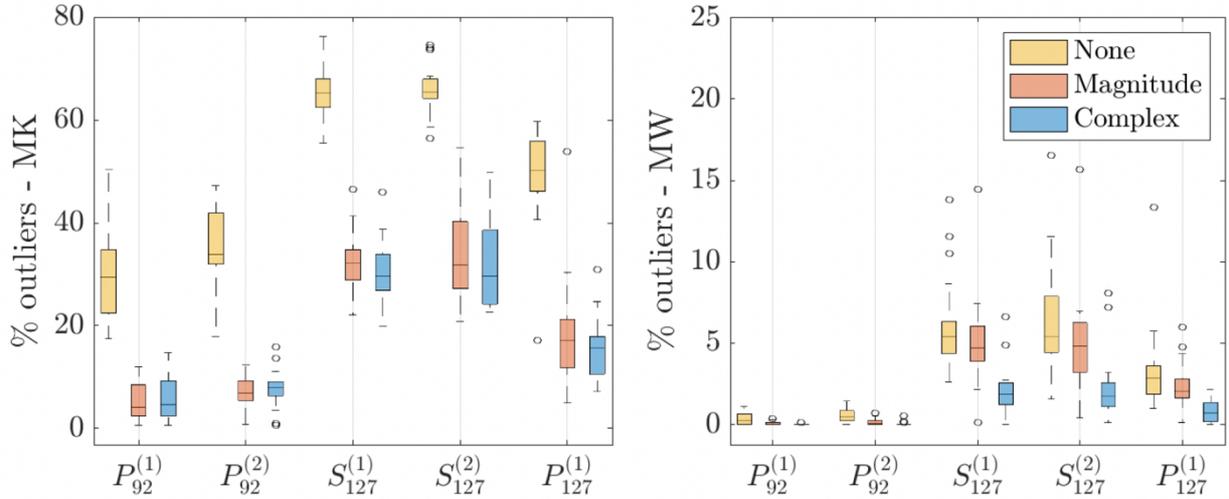

Figure 3: Percentage of outliers in MK and MW with each denoising method in the PLIC. The number of K outliers reaches 70% on the Skyra system, while the number of W outliers is consistently less than 10 % on both scanners and the lowest after denoising.

Next, in voxel-wise analyses, variability of both *K* and *W* parameters are measured to provide a complete account for the effect of large outliers in *K* parameters. In ROI-wise comparisons, we examine only *K* parameters after thresholding outliers from regions of interest (thresholds chosen as −1 < *K* < 10).

*3.1. Voxel-wise Variability*

Voxel-wise test-retest CV maps averaged over subjects are shown in Figure 4, for MD, FA, MW, AW, and RW (corresponding voxel-wise maps for kurtosis are provided in supplementary material Figure S.1). Voxel-wise CV is increased in regions near tissue boundaries due to the effect of noise limiting the precision of registration across subjects. However, this effect is reduced when using denoised dMRI, as visible in the MW-map around the genu (Figure 4).

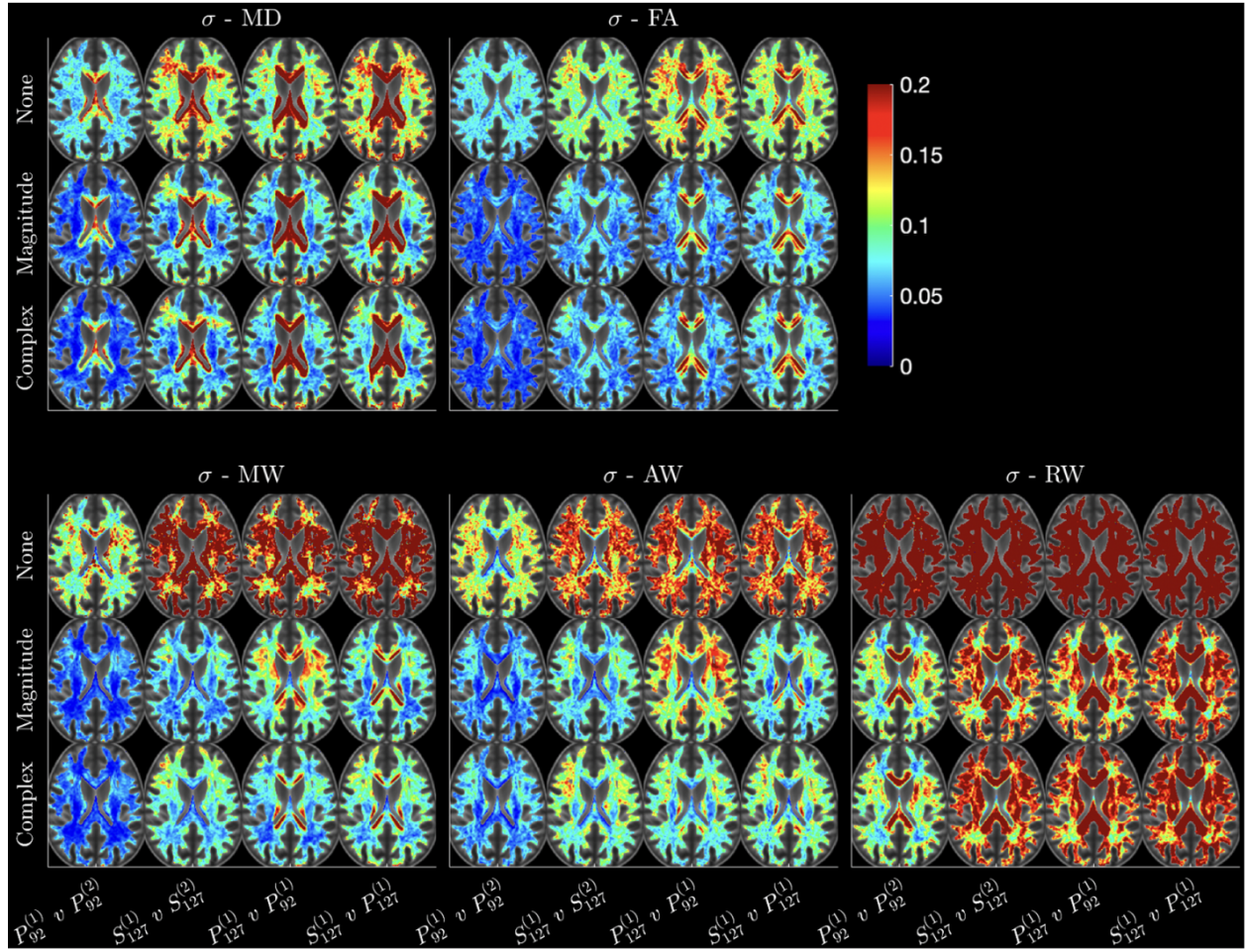

Figure 4: Maps of coefficient of variation in WM without normalization (effect size = σ) averaged over all 20 subjects. CV maps were computed in each subject's space and then warped into a common space prior to averaging. For each parameter left two columns show within-scan repeatability ($P_{92}^{(1)}$ vs $P_{92}^{(2)}$ and $S_{127}^{(1)}$ vs $S_{127}^{(2)}$), third column shows cross-protocol variability ($P_{92}^{(1)}$ vs $P_{127}^{(1)}$), and fourth column shows cross-scanner variability ($P_{127}^{(1)}$ vs $S_{127}^{(1)}$).

To minimize the effect of misregistration, Table 2 lists the mean voxel-wise CV pooled across different subjects within the splenium, posterior limb of the internal capsule (PLIC), anterior corona radiata (ACR), and thalamus, respectively. Regional CVs for kurtosis (K instead of W) including outliers are also shown in supplemental Figure S.2. CVs for MK and RK often exceed 20% even after denoising due to the higher probability for outliers. W parameters offer better robustness to outliers, such that without denoising, MW in the internal capsule has voxel-wise CV on the order of 9-12% and denoising lowers test-retest variability on the Prisma system down to 3 − 4%. FA has variability on the order of 7-10% without denoising in highly aligned WM regions, and denoising lowers variability to 4-6% for both Prisma and Skyra data.

|  |  | PRISMA$^{(1)}_{92}$ PRISMA$^{(2)}_{92}$ | | | SKYRA$^{(1)}_{127}$ SKYRA$^{(2)}_{127}$ | | | PRISMA$^{(1)}_{92}$ PRISMA$^{(1)}_{127}$ | | | PRISMA$^{(1)}_{127}$ SKYRA$^{(1)}_{127}$ | | |
|---|---|---|---|---|---|---|---|---|---|---|---|---|---|
|  |  | Complex | Magnitude | None | Complex | Magnitude | None | Complex | Magnitude | None | Complex | Magnitude | None |
| SPLENIUM (CC) | MD | 0.056 | 0.060 | 0.089 | 0.103 | 0.102 | 0.158 | 0.174 | 0.148 | 0.180 | 0.224 | 0.197 | 0.241 |
|  | AD | 0.045 | 0.048 | 0.097 | 0.078 | 0.075 | 0.149 | 0.115 | 0.102 | 0.165 | 0.143 | 0.130 | 0.199 |
|  | RD | 0.149 | 0.157 | 0.218 | 0.291 | 0.289 | 0.374 | 0.382 | 0.327 | 0.392 | 0.515 | 0.451 | 0.495 |
|  | FA | 0.051 | 0.052 | 0.079 | 0.075 | 0.076 | 0.102 | 0.130 | 0.110 | 0.138 | 0.159 | 0.148 | 0.167 |
|  | MW | 0.042 | 0.041 | 0.093 | 0.105 | 0.084 | 0.196 | 0.104 | 0.149 | 0.173 | 0.121 | 0.116 | 0.218 |
|  | AW | 0.065 | 0.061 | 0.106 | 0.094 | 0.071 | 0.141 | 0.102 | 0.193 | 0.187 | 0.111 | 0.112 | 0.172 |
|  | RW | 0.123 | 0.148 | 0.429 | 0.308 | 0.254 | 0.909 | 0.248 | 0.298 | 1.240 | 0.282 | 0.303 | 1.084 |
|  | MK | 0.077 | 0.085 | 1.199 | 0.269 | 0.368 | 4.584 | 0.311 | 0.384 | 3.230 | 0.343 | 0.295 | 4.793 |
|  | AK | 0.065 | 0.061 | 0.106 | 0.094 | 0.071 | 0.140 | 0.102 | 0.193 | 0.186 | 0.111 | 0.112 | 0.172 |
|  | RK | 0.089 | 0.112 | 1.958 | 0.518 | 0.367 | 5.299 | 0.412 | 0.527 | 3.552 | 0.273 | 0.312 | 5.162 |
| PLIC | MD | 0.034 | 0.035 | 0.069 | 0.051 | 0.048 | 0.102 | 0.051 | 0.053 | 0.089 | 0.061 | 0.058 | 0.105 |
|  | AD | 0.039 | 0.039 | 0.092 | 0.051 | 0.053 | 0.132 | 0.058 | 0.062 | 0.117 | 0.058 | 0.060 | 0.135 |
|  | RD | 0.066 | 0.069 | 0.136 | 0.115 | 0.115 | 0.214 | 0.102 | 0.104 | 0.183 | 0.142 | 0.144 | 0.227 |
|  | FA | 0.045 | 0.046 | 0.092 | 0.068 | 0.071 | 0.116 | 0.068 | 0.070 | 0.120 | 0.082 | 0.086 | 0.127 |
|  | MW | 0.026 | 0.027 | 0.070 | 0.045 | 0.039 | 0.116 | 0.042 | 0.091 | 0.110 | 0.051 | 0.042 | 0.115 |
|  | AW | 0.054 | 0.047 | 0.111 | 0.069 | 0.057 | 0.139 | 0.072 | 0.130 | 0.162 | 0.076 | 0.065 | 0.140 |
|  | RW | 0.058 | 0.060 | 0.221 | 0.101 | 0.091 | 0.519 | 0.081 | 0.092 | 0.405 | 0.113 | 0.102 | 0.547 |
|  | MK | 0.030 | 0.031 | 0.399 | 0.033 | 0.030 | 3.245 | 0.027 | 0.043 | 3.630 | 0.040 | 0.038 | 3.674 |
|  | AK | 0.054 | 0.047 | 0.111 | 0.069 | 0.058 | 0.139 | 0.072 | 0.130 | 0.161 | 0.077 | 0.065 | 0.139 |
|  | RK | 0.049 | 0.051 | 0.788 | 0.039 | 0.040 | 5.146 | 0.034 | 0.039 | 5.048 | 0.051 | 0.052 | 5.121 |
| ACR | MD | 0.039 | 0.041 | 0.074 | 0.069 | 0.067 | 0.121 | 0.063 | 0.064 | 0.100 | 0.079 | 0.074 | 0.127 |
|  | AD | 0.045 | 0.045 | 0.098 | 0.072 | 0.065 | 0.145 | 0.072 | 0.071 | 0.126 | 0.076 | 0.070 | 0.147 |
|  | RD | 0.062 | 0.064 | 0.112 | 0.107 | 0.104 | 0.177 | 0.089 | 0.091 | 0.162 | 0.122 | 0.116 | 0.191 |
|  | FA | 0.081 | 0.083 | 0.152 | 0.116 | 0.115 | 0.198 | 0.113 | 0.109 | 0.223 | 0.130 | 0.130 | 0.197 |
|  | MW | 0.031 | 0.029 | 0.074 | 0.065 | 0.050 | 0.128 | 0.058 | 0.113 | 0.133 | 0.070 | 0.059 | 0.131 |
|  | AW | 0.053 | 0.045 | 0.107 | 0.081 | 0.058 | 0.136 | 0.079 | 0.158 | 0.167 | 0.089 | 0.064 | 0.141 |
|  | RW | 0.059 | 0.057 | 0.188 | 0.115 | 0.092 | 0.421 | 0.103 | 0.086 | 0.364 | 0.119 | 0.100 | 0.450 |
|  | MK | 0.036 | 0.034 | 0.163 | 0.048 | 0.036 | 1.673 | 0.048 | 0.067 | 2.511 | 0.052 | 0.040 | 2.168 |
|  | AK | 0.053 | 0.045 | 0.107 | 0.081 | 0.058 | 0.136 | 0.079 | 0.158 | 0.167 | 0.089 | 0.065 | 0.141 |
|  | RK | 0.057 | 0.056 | 0.277 | 0.055 | 0.044 | 3.128 | 0.057 | 0.041 | 4.512 | 0.058 | 0.046 | 3.823 |
| THALAMUS | MD | 0.058 | 0.060 | 0.092 | 0.088 | 0.084 | 0.146 | 0.116 | 0.117 | 0.161 | 0.129 | 0.123 | 0.182 |
|  | AD | 0.059 | 0.060 | 0.116 | 0.083 | 0.076 | 0.167 | 0.110 | 0.105 | 0.207 | 0.116 | 0.107 | 0.196 |
|  | RD | 0.074 | 0.077 | 0.117 | 0.121 | 0.114 | 0.191 | 0.149 | 0.150 | 0.226 | 0.176 | 0.167 | 0.239 |
|  | FA | 0.113 | 0.119 | 0.236 | 0.151 | 0.144 | 0.260 | 0.161 | 0.163 | 0.389 | 0.175 | 0.168 | 0.300 |
|  | MW | 0.055 | 0.052 | 0.132 | 0.100 | 0.086 | 0.221 | 0.112 | 0.182 | 0.218 | 0.115 | 0.117 | 0.238 |
|  | AW | 0.073 | 0.064 | 0.128 | 0.109 | 0.088 | 0.151 | 0.108 | 0.207 | 0.185 | 0.123 | 0.117 | 0.178 |
|  | RW | 0.092 | 0.092 | 0.296 | 0.142 | 0.124 | 1.116 | 0.186 | 0.157 | 0.622 | 0.179 | 0.158 | 0.884 |
|  | MK | 0.055 | 0.054 | 0.406 | 0.050 | 0.045 | 3.454 | 0.074 | 0.096 | 3.870 | 0.069 | 0.068 | 2.729 |
|  | AK | 0.073 | 0.064 | 0.128 | 0.110 | 0.088 | 0.150 | 0.109 | 0.208 | 0.183 | 0.124 | 0.118 | 0.176 |
|  | RK | 0.074 | 0.074 | 0.758 | 0.048 | 0.047 | 3.958 | 0.081 | 0.066 | 4.915 | 0.075 | 0.068 | 3.368 |

Table 2: Mean coefficients of variation for each DTI and DKI parameter and denoising type evaluated in five regions: Splenium of the corpus callosum, posterior limb of the internal capsule (PLIC), anterior limb of the internal capsule (ALIC), and thalamus. We show coefficients of variation for each of the three comparisons: within-scan variability ($P^{(1)}_{92}$ vs $P^{(2)}_{92}$ and $S^{(1)}_{127}$ vs $S^{(2)}_{127}$), Cross-protocol variability ($P^{(1)}_{92}$ vs $P^{(1)}_{127}$), and cross-scanner variability ($P^{(1)}_{127}$ vs $S^{(1)}_{127}$).

The largest benefit of denoising and lowering the noise floor presents when comparing different echo times (Table 2). In AW we observe decreases in CV from 16.2% to 7.2% in the PLIC with complex denoising, along with corresponding large reductions in CV in other remaining WM areas. In line with the maps shown in Figure 4, we observe that CVs are consistently the lowest in areas such as the internal capsule, that are less prone to partial volume effects and Gibbs ringing. Indeed, regions near the ventricles such as the corpus callosum often have relatively higher variability.

Box plots of mean voxel-wise CVs in PLIC are shown in Figure 5a over all subjects (corresponding boxplots for kurtosis CVs are provided in supplemental material). We observe strongest effects of denoising in $W$ between data with differing echo times ($P_{92}^{(1)}$ vs $P_{127}^{(1)}$), where denoising complex data leads to a 12% decrease in CV in AW, along with marked improvements in CV in AD and FA from $\sim$ 9% down to 6%. Denoising (both magnitude and complex) give strong improvements in voxel-wise CV in kurtosis parameters because denoising helps minimize outliers in both $K$ and $W$ maps. While complex and magnitude denoising often give very similar changes in CV, MP complex gives the greatest improvement in cross-echo time reproducibility (3-7% in MD, MW, and FA) because of the reduction in noise floor.

Figure 5b and 5c show the results of concordance analysis on voxel-wise data. It is visible from concordance correlation coefficients (shown in Fig. 5b) that denoising leads to stronger correlations and lower variance in all test-retest datasets. In fact, without denoising, voxel-wise correlations drop to as low as 0.13 across scanners and protocols in kurtosis values. Correlation and Bland-Altman analyses demonstrate the strongest correlations and narrowest voxel-wise parameter distributions between scans on the same scanner with the same echo-time after denoising. The correlations are lower for data acquired on different scanners (same echo time), or with different echo time (same scanner). Bland-Altman plots (Fig. 5c) show the shift in parameter values resulting from decreased noise floor with complex denoising, particularly in MW. The middle row of Fig. 5c shows how denoising complex data decreases WM kurtosis values while also minimizing test-retest error across all comparisons.

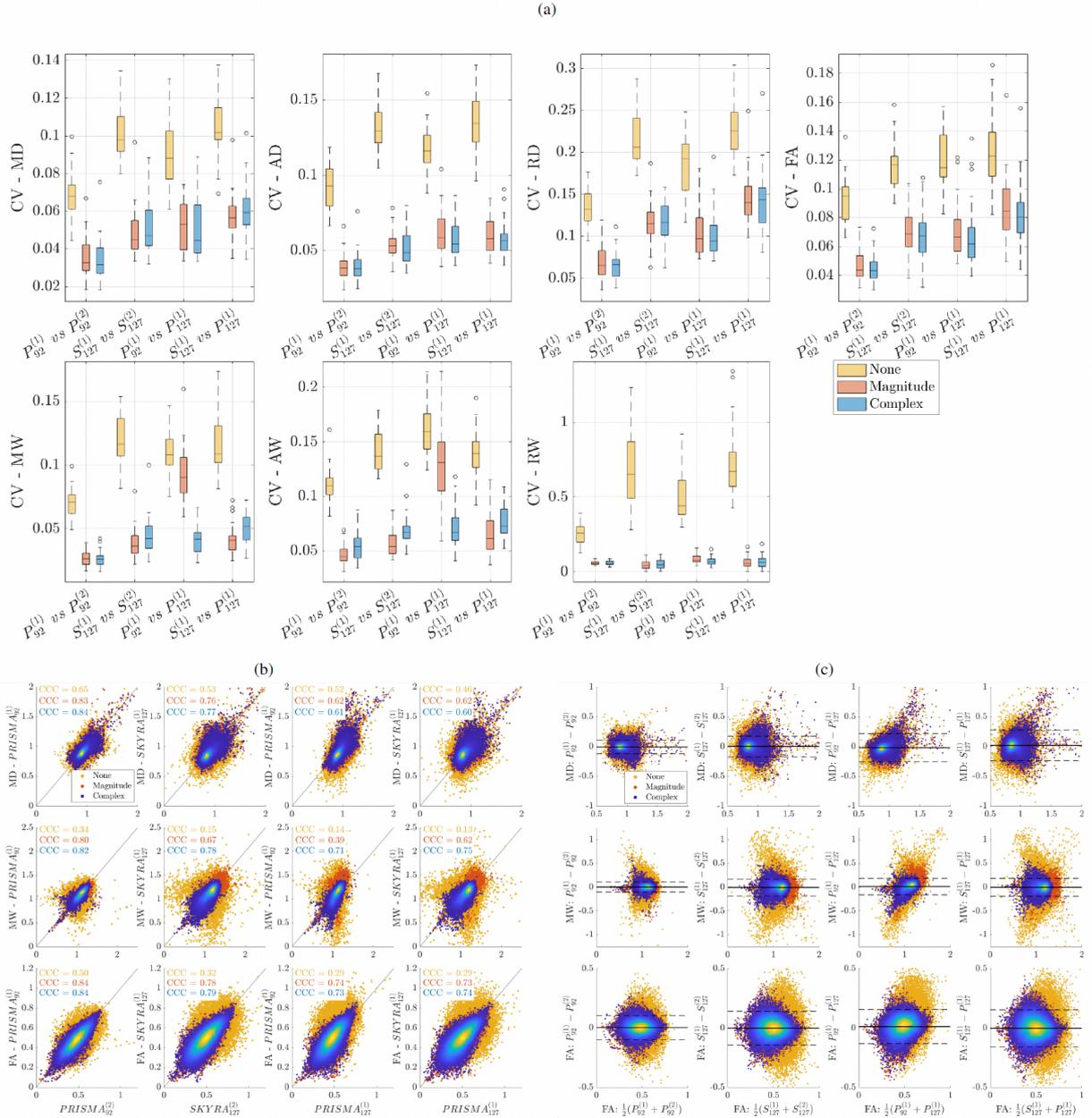

Figure 5: (a) Box plots for 20 subjects where CV maps were pooled over ROIs. Plots show $\sigma_x/\mu_x$ after pooling DTI and DKI parameters over whiter matter (PLIC, ALIC) over all subjects. X-axes show each of the four comparisons: within-scan variability ($P_{92}^{(1)}$ vs $P_{92}^{(2)}$ and $S_{127}^{(1)}$ vs $S_{127}^{(2)}$), Cross-protocol variability ($P_{92}^{(1)}$ vs $P_{127}^{(1)}$), and cross-scanner variability ($P_{127}^{(1)}$ vs $S_{127}^{(1)}$). Note that CVs here are not subject to non-linear warps, as they are pooled from subject-wise templates rather than the population template. (b) Voxel-wise scatter plots over white matter for all 20 subjects. Concordance Correlation Coefficients (CCC) are shown on the upper left corner of each plot for DTI and DKI parameters over white matter (internal capsule). (c) Bland-Altman plots over WM for all 20 subjects. Y axes show the error between test-retest measurements and X axes show the mean. Solid black line shows an error of 0, and dashed lines indicate error greater than 3 standard deviations away from the mean.

*3.2. ROI-wise Variability*

Figure 6 shows averaged DTI and DKI parameters over ROIs and the corresponding CVs rather than voxel-wise CVs. Figure 6a shows boxplots of mean DTI and DKI parameters respectively in WM over all 20 subjects, for all 5 scans, and for all denoising methods. While there is no ground truth, we know that repulsion due to noise causes artificial increases to FA, AD, and decreases in RD[65,66]. After either magnitude or complex denoising we observe a consistent decrease in bias across all diffusion tensor parameters. Furthermore, only after complex denoising, AW values, and to a lesser extent in MW, are lowered due to the decreased effect of the noise floor.

There is less advantage to denoising when averaging over hundreds of voxels within a region as compared to voxel-wise analyses (as SNR increases with the square root of the number of voxels in the ROI), but CVs still improve by 1-9% (supplement Table S.1) in each diffusion parameter in the ACR after noise variance and noise floor reduction in cross-echo time data. Since magnitude and complex denoising appear to work similarly on the level of ROIs, this reduction in CV can likely be attributed to noise floor reduction. This noise floor reduction is particularly pronounced in $W$ parameters compared to DTI parameters, where the effect of the noise floor can generate up to 15% bias in AW and RW.

Figure 6b shows boxplots of ROI-wise CV DTI parameters and DKI parameters respectively in the PLIC over all 20 subjects, for the 4 comparisons, and for all 3 processing methods. Test-retest variability on Prisma and Skyra hover between 1-4% for diffusion and kurtosis values, and the lowest test-retest variability is observed on the Prisma due to its high SNR because of the shorter TE on this scanner. Cross-protocol and cross-scanner variability are higher due to inconsistency in echo time and hardware respectively, resulting in greater differences in SNR between compared datasets. A notable decrease in CV is observed in AW and MW when examining cross-protocol variability because of bias reduction through complex denoising - resulting in decreased eigenvalue repulsion and noise floor. Supplement Table S.1 shows mean values of ROI-averaged CV over all 20 subjects. We found average test-retest CV for MD on the Prisma scanner to be about 1-2% in WM, on the Skyra we found slightly larger CVs, but also on the order of 1-2% regardless of whether denoising was used. Cross-protocol CV was higher, in the regime of 3-4%, and cross-scan CV in the range of 2-3%. Complex denoising consistently lowers variability in diffusivity, however this benefit was quite small (0.5%-1.0% improvements) at the level of the ROI. The strongest reduction in variability is observed for AW and complex denoising thanks to noise floor reduction prior to ROI averaging.

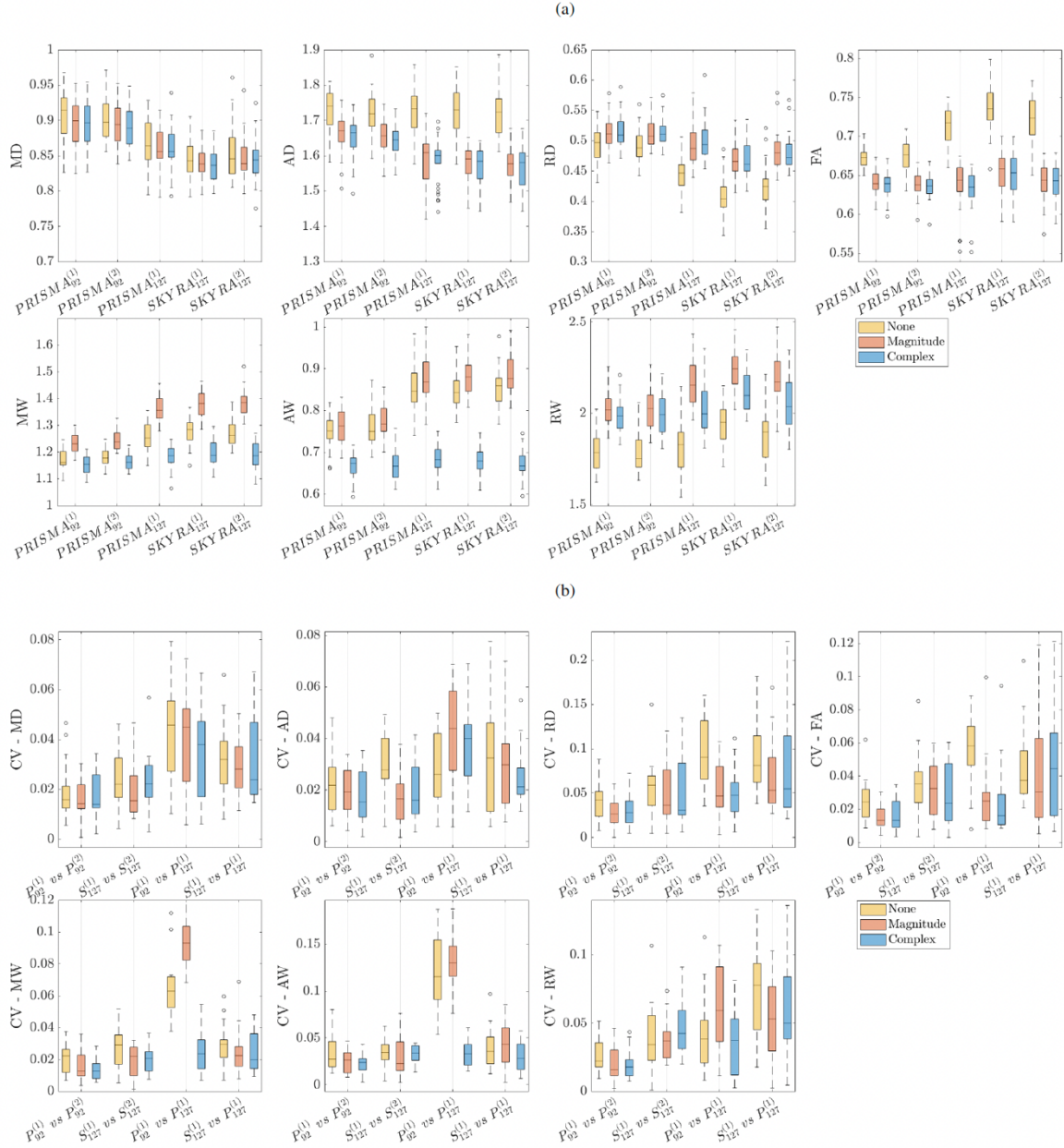

Figure 6: (a) Regional values over white matter averaged over 20 subjects. White matter regions shown: PLIC, ALIC, Genu of CC, Body of CC, Splenium of CC. DTI parameters are shown in the top row and DKI parameters in the bottom row. (b) ROI-wise coefficients of variation for 20 subjects. Plots show $\sigma_x/\mu_{x^x}$ after averaging DTI and DKI parameters over whiter matter (PLIC, ALIC). X-axes show each of the four comparisons: Cross-protocol variability ($P_{92}^{(1)}$ vs $P_{127}^{(1)}$), within-scan variability ($P_{92}^{(1)}$ vs $P_{92}^{(2)}$ and $S_{127}^{(1)}$ vs $S_{127}^{(2)}$), and cross-scanner variability ($P_{127}^{(1)}$ vs $S_{127}^{(1)}$).

3.4. *Comparison between harmonization and denoising*

RISH harmonization was applied to voxel-wise data. Figure 7 shows ICCs between each test-retest comparison and for data processed with denoising and/or harmonization. Across all comparisons ICC based on harmonization alone is lowest (with the exception of MD between scanners), corroborating the overall benefit of denoising prior to harmonization. In test-retest comparisons denoising along with RISH improved ICC from 0.63 to 0.93, and from 0.24 to 0.83 on Prisma and Skyra respectively, for FA. Denoising and harmonization combined greatly improved repeatability for MW, where either denoising alone or harmonization alone gave ICC values < 0.5, but after combined processing ICC values improved to 0.62 on the Prisma.

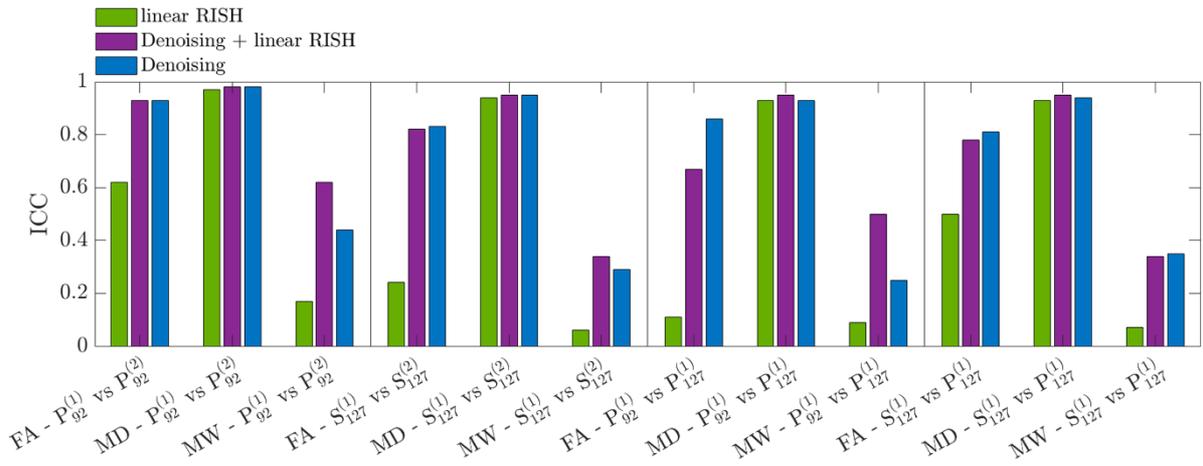

Figure 7: Intraclass correlation coefficients for over four comparisons: within-scan variability ($P_{92}^{(1)}$ vs $P_{92}^{(2)}$ and $S_{127}^{(1)}$ vs $S_{127}^{(2)}$), Cross-protocol variability ($P_{92}^{(1)}$ vs $P_{127}^{(1)}$), and cross-scanner variability ($P_{127}^{(1)}$ vs $S_{127}^{(1)}$). Denoising of complex data consistently improves the ICC.

*3.5. Noise Floor Estimation*

Figure 8 shows the noise floors for dMRI data denoised using MP-complex, MP-magnitude, and no denoising for each of the 5 scans. We found that the baseline noise level for non-denoised (or magnitude denoised) data is about 1.8-2.5% of the dynamic range of the DWI dataset. Complex denoising lowers the noise floor to 0.4-1%, on the Prisma system constituted a 2.5× reduction in noise floor. This reduction in noise floor propagates through tensor estimation and leads to the decreases in parameter bias present in Figure 6a.

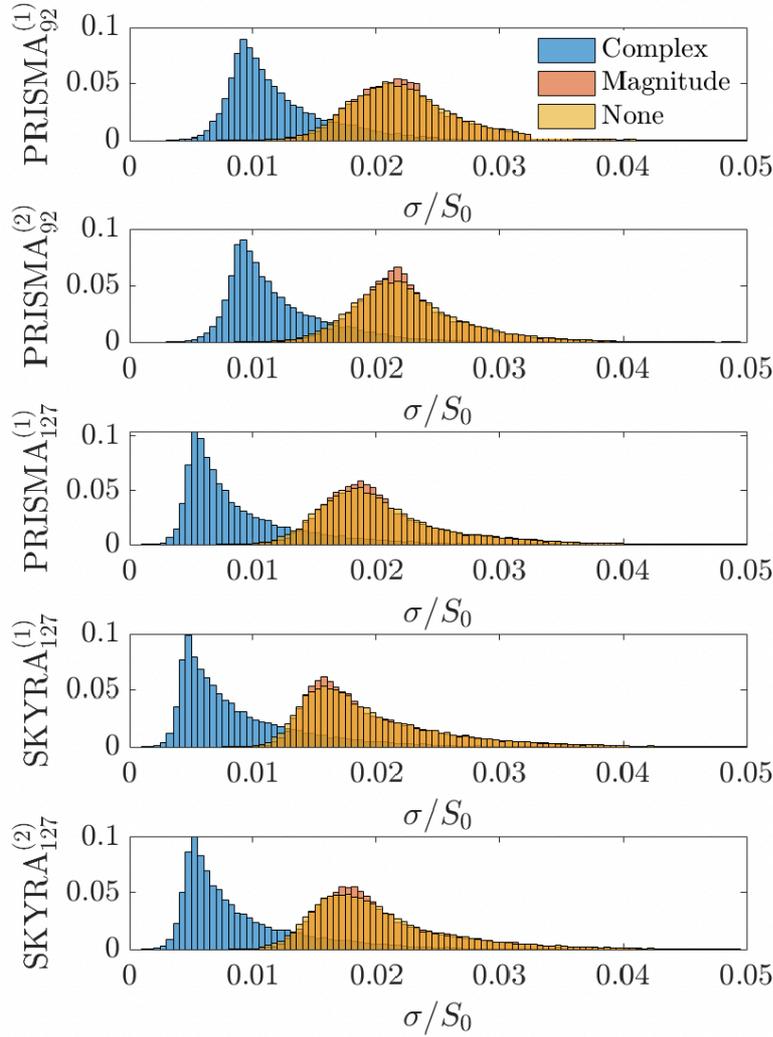

Figure 8: Noise floor estimation for each of the five scans according to Eq. (9). Distributions shown come from pooled voxels in the ventricles over all subjects. As expected, MP magnitude shows the same noise floor as no-denoising (as it does not remove the Rician bias). MP complex decreases the noise floor by a factor of 2 – 4.

*3.6. Statistical Power Estimation*

Figure 9 shows the results of a statistical estimate of the sample size required to detect 10% difference in group means, for groups with equal variance. This analysis was performed for a voxel located in a common WM region (PLIC). Here we found that denoising universally lowers the number of subjects required to reach statistical significance ($p = 0.05$ in a 2-sided t-test). On the Prisma scanner, denoising allowed for sample size decreases of 50%, 60%, and 62.5% for MD, MW, and FA respectively. On the Skyra we found that denoising allowed for corresponding decreases of 61.1%, 76.2%, and 40.9%.

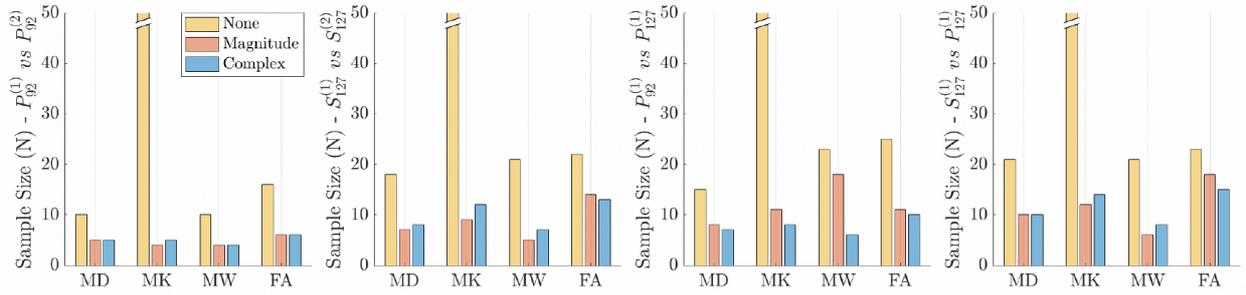

Figure 9: Required sample size needed to significantly distinguish groups separated by an effect size of 5% with α = 0.05 and a power of 0.8. Results are based on ROI means and standard deviations over all 20 subjects. Rows show each of the four comparisons within-scan variability ($P_{92}^{(1)}$ vs $P_{92}^{(2)}$ and $S_{127}^{(1)}$ vs $S_{127}^{(2)}$), Cross-protocol variability ($P_{92}^{(1)}$ vs $P_{127}^{(1)}$), and cross-scanner variability ($P_{127}^{(1)}$ vs $S_{127}^{(1)}$). The y-axis is clipped at 50 subjects, however sample size for MK without denoising required over 2000 subjects in all cases, when not accounting the for presence of kurtosis outliers.

For data acquired using protocols with different TE ($P_{92}^{(1)}$ vs $P_{127}^{(1)}$), complex denoising had the largest impact on sample size improvement because it offers a correction for differing noise floors. Complex denoising allowed for a sample size decrease of 53.3%, 73.9%, and 60% in MD, MW, and FA. On different scanners with the same TE, ($S_{127}^{(1)}$ vs $P_{127}^{(1)}$), both magnitude and complex denoising allows for a decrease in sample size of similar magnitude, hence we can conclude that these datasets contain similar noise floors.

Further, we observed that without denoising, outliers in MW drove up voxel-wise variances, inhibiting a valid comparison for this parameter. When simulating group comparisons across data with varying TE, magnitude denoising enabled statistical significance with N=18 subjects and complex denoising enabled significant comparisons with just 6 subjects.

4. Discussion

The goal of this study was to measure the reproducibility of higher-order diffusion MR metrics.

Denoising[27] of magnitude or complex diffusion MRI data, together with targeted artifact removal[51], improves reproducibility of higher-order diffusion parameters across scanners and protocols. Denoising reduces variations across test-retest datasets from ∼ 15−20% to ∼ 5−10% in kurtosis indices at the level of individual voxels. Outliers decrease sharply when switching from K to W maps, and by applying denoising. Subsequent measurements on different scanners and different TE (92ms, 127ms) were found to have voxel-wise precision varying from 3 − 15% for both DTI and DKI metrics after denoising. Notably, we found the greatest improvement in reproducibility across data with differing signal-to-noise ratios (from varying echo times) when applying MPPCA to complex-valued data, likely because of the ∼ 2.5× noise floor reduction offered by complex denoising.

When comparing denoising directly to harmonization using linear-RISH, denoising outperformed harmonization in reducing bias from varying noise floors. Moreover, combining denoising with harmonization in voxel-wise assessments improved intra-scanner test-retest intraclass correlation coefficients (ICCs) by 55% for FA, over harmonization alone. This highlights denoising's critical role in maximizing data precision and reliability in multi-site dMRI studies.

This works stands out compared to previous works due to the comprehensive nature of the multi-shell diffusion MRI dataset of 20 subjects, which allows for evaluating 3 separate types of reproducibility, along with analysis of how noise varies in repeated voxel-wise measurements. Previous studies aimed to evaluate harmonization schemes include denoising as an aspect of their harmonization[69] either directly, or because of neural network harmonization where denoising is implicit to the network due to clean supervised training data. The values reported here are largely in line with prior results of DTI repeatability found in the literature[70–73], where for ROIs in the corpus callosum, FA has been shown to have ∼ 2 − 3% within-scan and between-scan variability in aligned WM regions. Our reported values also agree with a previous report of intra-scanner ROI mean kurtosis repeatability of 1 − 4%[67]. Similarly, we also show correspondence with reported reproducibility[33] for tissue microstructure parameters. Such reproducibility may improve our ability to measure subtle changes in nervous tissue functional organization and complexity.

DKI repeatability so far has not been as thoroughly studied compared to DTI due to difficulty in generating clean kurtosis maps from generalizable protocols. The protocols used here were minimized regarding the number of gradient directions and $b$-shells needed to measure kurtosis in a clinically feasible time frame (∼7 minutes, Table 1). We demonstrate here relatively low precision to estimate the kurtosis tensor using the conventional weighted linear least squares estimator, with conventional kurtosis tensor parameters prone to "black voxels" particularly in the case of no denoising (see Figure 2). These outliers stem from noise propagation for the definition (eq. 6), where we divide by low directional diffusivities and cause kurtosis to blow up to unphysical levels. To address this confound, we included the alternative definition, $W$ (eq. 7), which we show in Figure 3 to be more robust.

We found that MPPCA denoising has a strong impact on voxel-wise data, lowering CVs in kurtosis from ∼

40% down to 5%. As shown in Figure 4, the effect is most noticeable in homogeneous brain regions. However, voxel-wise CV is nonuniform across WM due to different levels of tissue homogeneity in different regions, and CV is typically higher in regions with tissue boundaries such as the genu and splenium of the corpus callosum, owing largely to the effect of registration-based spatial interpolations along with small misregistration effects. Similarly, voxel-wise CV in GM (thalami) is higher than in homogenous WM owing to heterogeneous tissue composition.

On the other hand, there is less of a need for denoising when analyzing large groups of voxels (Figure 6b, Supplementary Table S1). Indeed, ROI analyses are more reproducible because of averaging many voxels compared to voxel-wise analyses[68], but may be also less specific and obscured due to spatially local effects. Conversely, voxel-wise analyses better capture biological variability, in which case denoising profoundly improves reproducibility. Indeed, voxel-wise, cross-protocol and cross-scanner CVs for non-denoised denoised diffusion parameters reach up to 20%, and after denoising variability drops to 3-8% in most areas (Table 2). DTI differences between groups for varying psychiatric disorders can be as low as 1% and up to 50% effect size[63], therefore our observed reduction in CV (Figure 5A) gained through denoising may be clinically useful.

We also found that denoising decreases the sample size needed to reach statistical significance between groups separated by a 10% effect size by about 50% in MD, MW, and FA (Figure 9). Because of the intrinsic low SNR of our diffusion protocol (due to small voxel size and short scan time), outliers in MK are quite prevalent when no denoising is applied, making it impossible to distinguish voxel-wise group differences without denoising. Remarkably, both magnitude and complex denoising reduced the required sample size to a reasonable level (5-15 subjects). Both complex and magnitude denoising produced similar effects, with the notable exception that when comparing groups who underwent scans with differing TE, where for MW, complex denoising provided an addition 66.6% reduction in sample size compared to magnitude denoising.

The additional benefit of MPPCA denoising of complex dMRI data (magnitude and phase) compared to denoising magnitude dMRI data is most pronounced in reducing parameter bias of higher-order dMRI metrics such as AK and to a lesser effect MK, as shown in Figure 5 and Figure 6a, particularly lowering the variability between protocols with different TE. Interestingly, the deviation of the regression slope from 1 in cross echo time (92 vs 127ms) CV of MW data (Figure 5b) is reduced most after complex denoising due to lowering the noise floor caused by higher noise floor in the long-TE scan.

Complex MP-denoising has the potential for further improvement since the algorithm uses a "two-pass" process, where we run the MP algorithm twice, first to estimate the noise free signal phase and then to perform the actual denoising. For this method to work effectively, these two steps must be sufficiently independent, i.e., the first step should not introduce noise correlations into the second step. If the patches used in each step are too similar, the denoising will not perform as robustly in the second denoising step. In this study we tried to make the patches as different as possible (15 × 15 2d patches during phase estimation, 3d adaptive patches during the actual denoising), however in theory the information in each patch should be perfectly orthogonal to maximize the performance of this approach. This is the reason why in some cases ($P_{127}^{(1)}$ vs $S_{127}^{(1)}$ AW in Fig. 5a) we observe slightly better denoising performance in MP-magnitude compared to MP-complex.

Here we show that the MPPCA denoising approach provides not only a notable and quantifiable improvement in reproducibility of cross-protocol dMRI parameters, but also provides a powerful tool for data harmonization. Indeed, due to the different echo times and numbers of gradient directions used in $P_{92}^{(1)}$ and $P_{127}^{(1)}$ protocols, the resulting dMRI data have differing SNR and noise floor (see figure 8). Without denoising, this results in parametric maps with differing levels of bias, and different numbers of outliers in kurtosis parameters owing to the different levels of precision in the raw data. Hence, reducing the noise floor by denoising complex dMRI data has the potential to improve parameter accuracy and *harmonize* data from sources with different SNR and thus, different noise floor. This is exemplified by the results of the harmonization analysis (Figure 7). We found that the impact of both complex denoising and harmonization was greatest when adjusting for the noise floor induced bias in FA in data with differing echo times. Since complex denoising is able to accurately reduce noise floor bias, it should be included as an essential first step to harmonizing data from separate sources.

Limitations of this study include the lack of a ground truth inhibiting us from knowing the exact degree of bias and variance induced by protocol specific effects. In addition, there are also additional sources of test-retest variability that were not measured here, including differing vendors, field strengths, q-space sampling regimes, and number of head coils, but could be the subject of future work. We assessed only young adult healthy controls and it should be noted that age or pathological changes to tissue microstructure may also increase diffusion parameter variability.

MP complex code is available as part of the DESIGNER-v2 diffusion MRI processing pipeline (https://github.com/NYU-DiffusionMRI/DESIGNER-v2) and can be run either through python or through a dedicated Docker container.

5. Conclusion

MPPCA denoising[27] reduces variability across scanners and echo times to 3-5% at the ROI-level and 5-10% voxel-wise for DTI and DKI metrics, and minimizes the sample size required for population-wise statistics by 40–70%. Denoising of complex dMRI enables noise floor reduction by up to 60%, and enables harmonization across scanners and protocols. Such improvement - that measurements from *individual voxels* become quantitative and reproducible- is an essential step towards bringing quantitative MRI and tissue microstructure imaging with MRI to clinical practice.